\documentclass{elsart}
\usepackage{epsfig,graphicx}
\def\DAF{DA$\Phi$NE}
\def\epem{e$^{+}$e$^{-}$}
\begin{document}
\begin{frontmatter}
\title{The role of KLOE and KLOE-2 in the search for a secluded 
gauge sector}
\author[Frascati]{Fabio Bossi}

\address[Frascati]{Laboratori Nazionali di Frascati dell'INFN, Frascati, Italy.}

\begin{center}
{\it Talk Given at the KLOE-2 Physics Workshop

Frascati April 9-10, 2009}
\end{center}

\begin{abstract}
The hypothesis of the existence of an hidden or secluded gauge sector 
with manifestations at
low or intermediate energies is motivated by several different recent 
astrophysical observations. 
At low and medium energy \epem\ colliders this sector gives clear signatures 
with cross sections that can be as large as 1 pb at 1 GeV.
Some of these signatures are straightforward, and can be realtively
easily isolated against background. Therefore,  
KLOE, with its collected 2.5 fb$^{-1}$, and KLOE-2 with its foreseen
detector's upgrades and larger data sample  are and will be able to test 
these models in deep detail. 
\end{abstract}

\end{frontmatter}

\section{Introduction}
In recent years, several astrophysical observations
have failed to find easy interpretations in terms of standard 
astrophysical and/or particle physics sources. 
A non exahustive lists of these observations includes 
the 511 keV photon signal from the galactic center observed by 
the INTEGRAL satellite~\cite{ref:int}, the excess in the 
cosimic ray positrons reported by PAMELA~\cite{ref:pam}, the total
electron and positron flux measured by ATIC~\cite{ref:atic} and 
the annual modulation of the DAMA/LIBRA signal~\cite{ref:dama}.     

An intriguing feature of these observations is the fact that 
all of them can be interpreted in terms of the existence of 
WIMP Dark Matter (DM) candidates belonging to a secluded gauge sector, 
under which the Standard Model (SM) particles are uncharged,
but can still weakly couple through a kinetic mixing mechanism
with typical mixing parameters $k$ naturally of the order 
10$^{-2}$-10$^{-3}$. More interestingly, from the point of 
view of the present note, the typical mass scale of the 
vector bosons related to the secluded  symmetry is of 
order 1 GeV.

A dramatic consequence of the above hypotheses is that they must 
induce observable effects in medium energy \epem\ colliders, 
such as the existing or future B-factories and the Frascati 
$\phi$-factory \DAF.
A rather comprehensive discussion of these effects can be found 
in references~\cite{ref:bpr} and~\cite{ref:est}. Although the 
two papers differ slightly in some detail of the underlying 
theoretical framework, the conclusion of the two are 
essentially the same. 

In the present note, I will move from the phenomelogical basis
suggested by these two articles, to discuss in a little detail 
the actual potentials of the KLOE experiment~\cite{ref:kloe}
 at \DAF, and of 
its proposed continuation, KLOE-2~\cite{ref:kloe2}. The paper is organized as 
follows. In the next section, a very qualitative description of
the physics models inspiring it is given. This description is not
meant to be fully comprehensive, neither in the details of the single model
nor in the variety of them, so I refer to the bibliography
reported therein, for a more complete panorama. 
In the following section, 
a short description of the KLOE apparatus and of its main performance
is given. Also, the plans for its usage in the near future are reported. 
Then, I concentrate on three possible signatures of this new physics at KLOE, 
namely the vector boson plus $\gamma$ signal, the dilepton plus missing energy, 
and the 
multilepton events. Since the physics potentials of KLOE, at least for 
some of these signals are already discussed in other papers, I  concentrate 
 mostly on the main experimental issues for all of them. 
Although this discussion is only semi-quantitative, we 
will see that one can safely conclude  
that there is an ample margin for KLOE and KLOE-2 to appreciably 
contribute to the field, by testing (some of) these models for a large 
part of the allowed free-parameters space.
It is also important to stress that the searches at KLOE and at the 
B-factories are largely complementary, so that it would be desirable 
to set-up some kind of networking among the various experiments, as well 
as between experiments and theorists in the field, to enhance the 
efficiency of the analysis process. 
  
\section{Main features of the secluded sector}
As stated above, there exist several different models 
with different gauge structures, which are interesting for the present 
discussion~\cite{ref:pos}~\cite{ref:aah}~\cite{ref:abpw}.
 A common feature of all of them is 
the existence of at least  
 a vector boson, the ``$U$'' from now on, of mass around 1
 GeV, responsible for mediating a new abelian U(1)$_{S}$ interaction. 
Together with it, it is also natural to conceive the existence of 
an elementary Higgs-like boson, the $h'$, which spontaneously breaks
the symmetry. The existence of the $h'$ is not strictly required by the  
observational inputs mentioned in the previous section. 
In fact the authors of reference 
\cite{ref:est} do consider also a model whithout it. However, the existence   
of the $h'$ is on the one hand rather natural theoretically, and on the 
other it has important consequences in terms of observable phenomenology, 
as it will be seen later on. 
The mass hierarchy between these two particles is not constrained. 
However it has important consequences in terms of possible decay channels,
so that one has to consider different detection strategies for the 
two cases $m_{h'} < m_{U}$ and $m_{h'} > m_{U}$.

Another ansatz that has relevant consequences in what follows, 
which is particularly stressed by the authors of reference~\cite{ref:bpr}, 
is that the mass of the $U$ boson satisfyes the relation 
$m_{U} << m_{WIMP}$, m$_{WIMP}$ being the mass of the DM weakly 
interacting candidate; 
this relation naturally leads to an enhanced WIMP 
annihilation cross section in the galaxy, as required by the data, 
and, more importantly for the present discussion, to an enhancement or
dominance of the leptonic branching ratios in the annhilation products.       

Both the $U$ and the $h'$ can be produced at \DAF, provided 
that their masses are less than m$_{\phi}$.
This is somewhat a disadvantage with respect to the B-factories, since 
the available phase-space is lower. Also, the luminosity of \DAF\ is 
about two orders of magnitude lower than the one of PEP-II and KEK-B.   
However there are two main advantages 
with respect to the B factories. On the one hand, the production cross 
sections scale as 1/s, which essentially compensates the lower luminosity
of \DAF. On the other hand, \DAF\ is better suited for detecting particles 
with masses lower than 1 GeV, since the main physical backgrounds are 
typically due        
to QED radiative processes, whose cross sections increase with the 
distance between the 'new particle' mass and the collision energy.

\section{KLOE and KLOE-2}
The KLOE detector consists mainly of a cylindrical 
drift chamber (DC) with momentum resolution $\sigma_{p_t}/p_t < $0.4$\%$, 
and a highly hermetic elctromagnetic calorimeter with energy
resolution of $\sigma_{E}/E=$5$\% /\sqrt{E(GeV)}$. The DC~\cite{ref:kdc}
has its first sensitive layer at a radius of 30 cm from the interaction 
point (IP); this fact, 
combined with a  B field value of 0.52 T, results
in an effective cut-off value for the transverse momentum  
of the detectable charged particles 
produced at the IP of $\sim$40 MeV/c.   
Photons and electrons of energy down to $\sim$20 MeV and polar angle 
in the range
20$^{\circ}< \theta <$160$^{\circ}$ can be detected with 
good efficiency by the calorimeter~\cite{ref:kcal}. 
Inefficiencies below the per 
mil level are obtained for energies above $\sim$100 MeV.

The trigger~\cite{ref:ktr} uses both calorimeter and DC information. 
In the first case,
two energy deposits above 50 MeV in the barrel or 150 MeV in the endcaps
are required. In the second, the trigger is produced by the count
of $\sim$100 wires accumulated over an interval of 1 $\mu$s. 

KLOE has so far acquired 2.5 fb$^{-1}$ of data at the $\phi$(1020) 
resonance peak. A further run is foreseen for the beginning of 
year 2010 with the goal of acquiring other $\sim$5 fb$^{-1}$, 
as a first part of the so called KLOE-2 program. 
Soon after a program for upgrading the detector is  planned. 
For the purpose of the present discussion, the main upgrades 
will consist in the insertion of a very light 
internal tracking device (the Inner Tracker, IT), 
which would allow detection of charged particles
down to a radius of $\sim$12 cm. Crystal calorimeters (CCALT),
 placed in front
of the low-$\beta$ focussing quadrupoles, will help detecting 
photons emitted in the very forward direction, down to 
$\theta$=9$^{\circ}$.  
The total integrated luminosity
for this second stage of operation should reach at least 20 fb$^{-1}$.

\section{$U\gamma$ events}
One interesting process to be studied is \epem\ $\rightarrow U\gamma$. 
It has the advantage of being independent of the existence and on 
the details of the Higgs sector. Also, its expected  cross section
can be as high as 0.1 pb at \DAF\ energies, as shown in 
figure 5 of reference~\cite{ref:est}. The on-shell boson can decay
into a lepton pair, giving rise to a $l^{+}l^{-}\gamma$ signal.
 
The most relevant physics background comes from the parent 
QED radiative process, 
which has a much higher cross section but can be rejected by cutting 
on the invariant mass of the lepton pair, as already discussed in 
reference~\cite{ref:bcd}. 
There is however a relevant instrumental background that has to be taken
into consideration for the electron channel,
 namely \epem\ $\rightarrow \gamma\gamma$ with 
subsequent conversion of one of the two photons on the beam pipe, 
with a probability of $\sim$2$\%$. This effect can be identified and 
rejected cutting on the reconstructed invariant mass and vertex of the pair
as done for instance in the analysis of~\cite{ref:eeepp}. 
For m$_{U}$ larger than $\sim$200 MeV, 
the rejection factor is around 10$^{2}$. Taking into account that 
the calorimeter energy resolution is $\sim$35 MeV for photons of 510 MeV, 
and that the cross section of the \epem\ $\rightarrow \gamma\gamma$ 
events is of a few hundreds nanobarns, one obtains that a reasonable
background rejection can be obtained only for m$_{U}\geq$500 MeV. 

The insertion of the IT can be rather beneficial in this case, since 
it would help in a better definition of the pair production vertex. 
A quantitative statement on this issue, however, needs the use of 
a detailed Monte Carlo simulation, which is at present unavailable.    

For the muon channel, the above mentioned background is not present. 
One has to take into account however, the physical process 
\epem\ $\rightarrow \pi\pi\gamma$, that is relevant, since 
$\pi$-$\mu$ separation in highly untrivial at \DAF\ energies. 

A further consideration has to be done, concerning the final detection
efficiency. Actually, the process under study, as well as its
backgrounds, has an angular distribution proportional to
$(1 + cos^{2}(\theta))/sin^{2}(\theta)$, which results in a limited 
geometrical acceptance, since most of the events are in the forward
direction. Again, the proposed modification for the second phase of 
KLOE-2 should be beneficial, since they increase acceptance for 
both charged tracks, thanks to the IT, and for photons, thanks 
to the CCALT. 

All considered, one can conclude that the $l^{+}l^{-}\gamma$ 
signature at KLOE has reasonable chances to be useful to explore the region
m$_{U}>$ 400-500 MeV, $k \sim$10$^{-2}$. Some improvement can be 
obtained with KLOE-2, thanks to the higher expected statistics and
to a better background rejection capability.  

A final note has to be made on the possibility that the $U$ boson decays
into two neutral long lived (or stable) particles, 
either DM WIMPs or neutrinos (as discussed for instance in~\cite{ref:bcd}). 
In this case the signal would be a single photon plus missing energy. 
This signal fails to satisfy the KLOE trigger conditions, 
so it cannot be in the present KLOE data set. Moreover, even assuming
the implementation of a dedicated trigger for the future, 
it would be affected by copious physical as well as machine backgrounds, 
that produce a single photon signal in the calorimeter at a much higher rate. 
Here, a key requirement is a very high energy resolution which would help
isolating the signal peak over a broad background. Unfortunately the 
KLOE calorimeter is not conceived for such an high resolution, so 
that the observation of this signature at KLOE is essentially hopeless.    

\section{The higgs'-strahlung process}
Assuming the existence of a higgs' boson, a particularly interesting
process from the experimental point of view is the higgs'-strahlung 
\epem\ $\rightarrow Uh'$, which can be observed 
at KLOE if m$_{U}$+m$_{h'}<$m$_{\phi}$.
As stated above, the signature of this process heavily depends on the
existing relation between m$_{U}$ and m$_{h'}$. In this paragraph
we assume that the $h'$ is lighter than the $U$ boson; in this case it 
turns out to be very long-lived (see~\cite{ref:bpr}), so that the signature 
of the process will be a lepton pair, generated by the $U$ boson decay, plus 
missing energy. 

There are several advantages for this type of signal. 
Firstly, there are no other physical processes with the same signature. 
The background due to QED $l^{+}l^{-}\gamma$  events with a 
 photon lost by the calorimeter, is suppressed by a relavant factor 
due to the high detection efficiency of this device. Moreover, this 
kind of background would give rise to a missing momentum equal to 
the missing energy, while in the case of the signal these two quantities 
will be sizeably different, due to the non-zero $h'$ mass. In this case, 
differently from the one discussed in the previous paragraph, 
the key ingredient is the very high resolution of the DC as compared to
the calorimeter one. A third advantage in terms of both background 
rejection and detection efficiency is that the angular distribution of
the process is proportional to sin$^{3}(\theta)$, which peaks at 
$\theta$=$\pi/2$. Finally, for a wide choice of m$_{U}$ and m$_{h'}$ 
the trigger efficiency should exceed 90$\%$.

The only physical process that can give rise to a dangerous background
at \DAF, is the process $\phi \rightarrow K^{0}_{S}K^{0}_{L}$ followed
by a $K^{0}_{S} \rightarrow \pi^{+}\pi^{-}$ decay and the $K^{0}_{L}$
flying through the apparatus without interacting. This decay chain is 
relevant only for the $U \rightarrow \mu^{+}\mu^{-}$ channel and its 
amount can be well calibrated by using the events in which the $K^{0}_{L}$
is observed in the apparatus. If this background turns out to be 
a problem, however, one can always take data at $\sqrt{s}<$2m$_{K}$,  
that can be easily done at \DAF, without loss of luminosity.       

As a side note to the present discussion, it can be added that 
KLOE is also particularly well suited for the observation of events 
with a displaced decay vertex of the $h'$. These events might happen 
for particular combinations of the parameters which can make the 
decay path of the $h'$ lower than a few meters. Actually the detector was
conceived to maximize the efficiency for the decays of the $K^{0}_{L}$'s, 
that have a mean free path of 3.5 m at \DAF. 
 
\section{Multilepton events}
In the case m$_{h'}>$ m$_{U}$, then, it more frequently decays to a pair 
of real or virtual $U$'s. In this case one can observe events with 6 leptons
in the final state, due to the higgs'-strahlung process, or 4 leptons and 
a photon, due to the  \epem\ $\rightarrow h'\gamma$ reaction. 

Albeit very spectacular, these kind of events suffer of the fact that 
at KLOE they have a relatively limited allowed phase-space, especially
for the muon channel. Also, the higher the multiplicity, the lower is the 
value of the mimimum transvere momentum for the chargerd particles, resulting
in a possible sizeable loss of acceptance, since one has to reconstruct
completely the events.  Clearly, the all-electron channel is privileged
at KLOE. In order to make more quantitative statements, however, a detailed
Monte Carlo simulation is needed. 

It is important to stress that multilepton events are foreseen also 
in more complex models, with repect to the one we have so far 
taken into consideration. 
For instance in models with a confined sector, such as the one of 
reference~\cite{ref:abpw} one has the presence of vector 'hidden' mesons
that can easily decay to leptons, and scalar ones that cannot, so that 
a possible resulting signal is a multilepton  plus missing energy one
\footnote{I thank N. Toro to have pointed me out this fact.}.
This model is very appealing since it reconciles well with the hypothesis that
the dark matter is done by scalar particles, as suggested by the authors
of reference~\cite{ref:bf}.  
Obviously, the detectability of such kind of signals is very  much dependent
on the details of the model, so that, once again, quantitative statements
can be made only after a dedicated Monte Carlo study has been performed. 

\section{Overview and conclusions}
The hypothesis of the existence of an 'hidden' or 'secluded' gauge sector 
with manifestations at
low or intermediate energies is motivated by several different recent 
astrophysical observations. 
This new physics can have escaped detection so far by particle physics 
experiments, due to its weak coupling to ordinary matter. 
Actually, the best limits on it are presently due to the measured value 
of the electron $g-2$, as dicussed for instance in 
references~\cite{ref:est}~\cite{ref:bcd}.
At low and medium energy \epem\ colliders this sector gives clear signatures 
with cross sections that can be as large as 1 pb at 1 GeV. Although 
this value is relatively high with respect to present day standards, 
it must be stressed that previous 
generation collider experiments at or around this energy, 
such as those at ACO, ADONE and VEPP, 
have collected statistics of a few inverse picobarns at maximum, so that 
they could not have been able to observe anything. 

KLOE at \DAF, with its collected 2.5 fb$^{-1}$, and KLOE-2 with its foreseen
detector's upgrades and larger data sample  are and will be able to test 
these models in deep detail. 
Since the typical masses of these hypothetical particles range 
between a few MeV up to  $\sim$10 GeV, searches must be performed at different
facilities. KLOE can be the front-runner in the range between a few hundreds
MeV to $\sim$1 GeV, while for higher 
mass values the present and future B-factories will play a crucial role. 

A common effort between theorists and experimentalists, in terms of 
production of Monte Carlo generators, discussion on measurement
strategies and interpretation of the data would be highly desirable.

\section*{Acknowledgments}
I would like to thank many people for their help in the preparation
of the present manuscript: M. Pospelov, 
R. Essig, P. Schuster and N. Toro for clarifications about many aspects  
of their important papers. G. Isidori and S. Pacetti for having provided me  
basic formulas to work out some order-of-magnitude calculations. 
My KLOE colleagues, and especially A. De Santis, 
P. Santangelo, A. Sciubba and R. Versaci
for many useful discussions.

\end{document}